\documentclass[twocolumn,showpacs,aps,prl]{revtex4}%
\usepackage{color}
\usepackage{amsmath}
\usepackage{amsfonts}
\usepackage{amssymb}
\usepackage{graphicx}
\usepackage{float}
\usepackage{subfigure}
\usepackage[colorlinks,breaklinks,  linkcolor=blue, anchorcolor=blue,citecolor=blue,
urlcolor=blue ]{hyperref}

\usepackage{txfonts}%
\usepackage{dcolumn}
\usepackage{bm}

\makeatletter

\newcommand{\Rmnum}[1]{\expandafter\@slowromancap\romannumeral #1@}
\makeatother
\begin{document}
\title{Aharonov-Bohm effect in monolayer black phosphorus (phosphorene) nanorings}
\author{Rui Zhang$^{1,2}$, Zhenhua Wu$^{3\dag}$, X. J. Li$^{4}$, Kai Chang$^{1,2}$\footnote{kchang@semi.ac.cn; $^{\dag}$wuzhenhua@ime.ac.cn }}
\affiliation{$^{1}$SKLSM, Institute of Semiconductors, Chinese Academy of Sciences, 100083 Beijing, China}
\affiliation{$^{2}$College of Materials Science and Opto-Electronic Technology, University of Chinese Academy of Sciences, 100049 Beijing, China}
\affiliation{$^{3}$Key Laboratory of Microelectronic Devices and Integrated Technology, Institute of Microelectronics, Chinese Academy of Sciences, 100029  Beijing, China}
\affiliation{$^{4}$College of Physics and Energy, Fujian Normal University, 350007 Fuzhou, China }

\begin{abstract}
  This work presents theoretical demonstration of Aharonov-Bohm (AB) effect in monolayer phosphorene nanorings (PNR). Atomistic
  quantum transport simulations of PNR are employed to investigate the impact of multiple
  modulation sources on the sample conductance. In presence of a perpendicular magnetic field, we find that the conductance of both armchair and zigzag PNR oscillate
  periodically in a low-energy window as a manifestation of the AB effect.
  Our numerical results have revealed a giant magnetoresistance (MR) in zigzag PNR (with a maximum magnitude approaching two thousand percent). It is attributed to the AB effect induced destructive interference phase in a wide energy range below the bottom of the second subband.
  We also demonstrate that PNR conductance is highly anisotropic, offering an additional way to modulate MR. The giant
  MR in PNR is maintained at room temperature in the presence of thermal broadening effect.
\end{abstract}
\pacs{73.21.-f, 78.67.-n, 75.75.-c, 81.07.-b}
\maketitle

\section{INTRODUCTION}
Phosphorene, the single- and few-layer form of black phosphorus (BP), has been successfully fabricated by researchers very recently\cite{Likai,LiuHan,BP7}.
It holds great promise for applications in electronics and optoelectronics because of its excellent mechanical, optical, thermoelectric, and electronic properties\cite{Buscema,Das,Likai,LiuHan,Luwanglin,Rodin,chen,Tayari,Yong-Lian,Likai1,BP9,BP13,TD2,BP7,ezawa}. BP is the most stable allotrope among the phosphorus group also including white, red, and violet phosphorus\cite{Keyes,Jamieson}. It consists of phosphorus atom layers coupled by weak van der Waals (vdWs) interlayer interactions. Bulk BP possesses a direct band gap , this direct gap increases  when the film thickness decreases from bulk to few layers and eventually monolayer via mechanical exfoliation. Due to its unique structure in two dimensional (2D) materials family, the band structure, electrical conductivity, thermal conductivity, and optical responses of phosphorene are highly anisotropic\cite{LiuHan,Rodin,Tran,Fei,ezawa} which is different to other widely studied 2D materials such as graphene, monolayer boron nitride (BN), silicene, and transition metal dichalcongenide (TMDCs).
\par
As a newly emerged member of the 2D crystal family, phosphorene ignited a surge of research activities in the physics,
chemistry, and materials communities because of its interesting unique physical properties and its potential application in the
future. Various properties of phosphorene have been investigated theoretically and experimentally, e.g., field transistor effect\cite{Likai,Buscema,BP7,BP14}, strain modification\cite{Rodin,Fei,BP13,PH4}, optoelectronics and electronics\cite{BP3,Rudenko,BP21,BP22,BP23,BP24,BP25,ruizhang,Xiaoying1,Xiaoying2}, transport properties\cite{BP2,LiuHan,TD1}, excitons\cite{Tran,BP20}, heterostructures and PN junctions\cite{BP12,BP18,PH2}, and a recent experimental demonstration of the crystalline anisotropy impacted phase coherent transport properties in BP field-effect transistor\cite{Hemsworth}. Another experiment carried out by Masih Das, et al.\cite{masih} indicates that it¡¯s possible to sculpture
phosphorus nanoribbons experimentally which provide a possibility to make a PNR as proposed in this paper.
\par
Aharonov-Bohm (AB) effect\cite{aharonov} is an important phenomenon in quantum physics which has aroused many attentions in the passed decades. The AB effect in graphene nanostructure
including graphene nanoribbons\cite{Mre}, nanotube\cite{Tian}, and graphene nanoring\cite{Nguyen,faria,Romanovsky,Yannouleas} have been investigated.
\begin{figure}[b]
  \centering
  \includegraphics[width=0.5\textwidth]{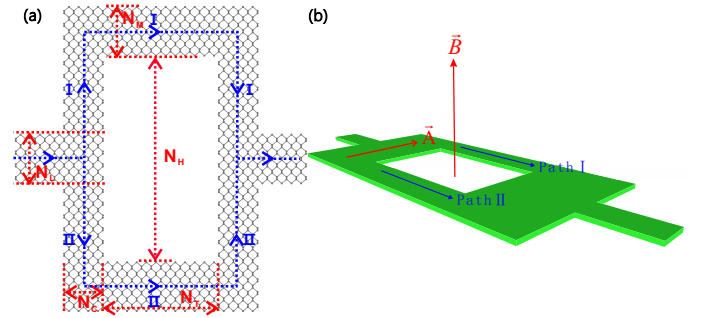}
  \caption{(Color online) Schematic diagram of a rectangular PNR subjected to a magnetic flux threading the PNR. (a) The parameters are in the unit of lattice constant \textbf{a} and \textbf{b}. In this example zigzag edged nanoring, $N_{L}=4$, $N_{M}=4$,  $N_{C}=4$,  $N_{T}=12$,  $N_{H}=16$. The parameters are in the unit of lattice constant \textbf{a} and \textbf{b}. (b) A uniform  magnetic field is applied perpendicular to the PNR. We use the Landau gauge $\vec{A}=(0,Bx,0)$. Electron transport through the central PNR via both path $I$, path $II$ and then recombine in the right lead.
  }\label{fig:1}
\end{figure}
\par
However, the AB effect in PNR remains unexplored, so in this work we theoretically investigate the transport properties of monolayer phosphorene nanorings (PNR) utilizing tight-binding (TB) method and recursive Green's function method. Transport properties of nanorings with different crystalline orientation, temperature, incident energy, magnetic filed are calculated. We find that the crystalline orientation of a nanoring giantly affects the quantum tunneling behavior and the value of magnetoresistance (MR), i.e., the MR is highly anisotropic in PNRs. Resonant tunneling can be obtained in both armchair and zigzag PNRs.
\par
The paper is organized as follows. In Sec. \Rmnum{2}, we present the TB model and the algorithm we calculate the transport properties of the system. In Sec. \Rmnum{3}, we briefly discuss the physics of AB effect and investigate the AB effect in PNRs. The MR of PNRs are demonstrated in Sec.\Rmnum{4}. Finally, we summarize our results in Sec. \Rmnum{5}.

\section{Model and Formulation}
In monolayer phosphorene each phosphorus atom is covalently bonded with three
adjacent phosphorus atoms to form a low puckered honeycomb structure. Phosphorene has an irregular honeycomb structure
with lattice constants  $a$=4.38{\AA} and $b$=3.31{\AA}. There are four phosphorus atoms in a unit cell.
The TB Hamiltonian for the PNRs can be written as\cite{Rudenko}
\begin{equation}\label{equ:1}
  H_{C}=\sum\limits_{i\neq
    j}t_{i,j}c_{i}^{\dagger }c_{j} ,
\end{equation}
where the summation runs over all the lattice sites of PNRs, $c_i^\dag$ ($c_j$) is the creation (annihilation)
operator of the electron at site $i$ ($j$), and ${t_{i,j}}$ are the hopping energies.
Five hopping links are needed to be taken into consideration\cite{Rudenko}. The related hopping integrals are
$t_{1}$=$-$1.220 eV, $t_{2}$=3.655 eV, $t_{3}$=$-$0.205 eV, $t_{4}$=$-$0.105 eV, and $t_{5}$=$-$0.055 eV.
The band gap of MLP given by this TB model is 1.52 eV with the valence band maximum (VBM) and a conduction band
minimum (CBM) located at $-$1.18 eV and 0.34 eV respectively\cite{Xiaoying2}.
When we consider a magnetic field $B$ applied perpendicularly to the plane of a PNR, the transfer integral
becomes ${\widetilde{t}_{i,j}}$ = ${t_{i,j}}{e^{i{\phi _{i,j}}}}$,
where $\phi_{i,j}=\frac{e}{\hbar}\int\nolimits_{r_{i}}^{r_{j}}d\mathbf{l\cdot A}$ is the Peierls phase. As we use Peierls substitution, which means
the magnetic field is not only applied on the hole in the nanoring but also applied on the lattice in the nanoring, which means the electrons in
the nanoring feels a field as $\vec{B}{\ne}0$,$\vec{A}{\ne}0$, the non-local part of AB effect mentioned in reference\cite{aharonov} is still reserved because of
the magnetic flux cross the big hole in the nanoring which can be felt by the electron in the nanoring too.
In our calculation, the magnetic field $\vec{B}$ is homogeneous, we take the Landau gauge, vector potential $\vec{A}=(0,Bx,0)$.
The magnetic flux $\phi=\frac{Bab}{2}$  through a plaquette is in unit of $\phi_{0}=\frac{\hbar}{e}$.
\par
The system is composed of a central mesoscopic conductor, i.e., the PNR and two semi-infinite leads. For a large PNR, the dimension of the Hamiltonian matrix $H_{C}$ is huge, and recursive Green's function algorithm is adapted in this work. We start by dividing
the system into vertical principal slices. The interaction only exists in/between adjacent slices. The Hamiltonian matrix block of the first slice, containing the self-energy of the left lead, is inverted and added to the block of the next slice to its right.
This procedure is repeated until we add the block of the last slice which contains the self-energy of the right lead. The conductance is associated with the scattering properties of the electron through the conductor region and is determined by the transmission probability via Landauer B\"{u}ttiker formula \cite{Landauer,Datta}
\begin{equation}
  \label{equ:2}
  G=\frac{2e^2}{h}T,
\end{equation}
where the conductance $G$ and transmission probability $T$ both depend on the incident energy $E_{f}$. In the following we adopt $G_{0}=\frac{2e^{2}}{h}$ as the unit of conductance. The transmission probability $T$ can be expressed in terms of the Green functions of the conductor and the coupling of the conductor to the leads ,
\begin{equation}\label{equ:3}
  T=Tr\left(  \Gamma_{L}G_{C}^{r}\Gamma_{R}G_{C}^{a}\right),
\end{equation}
where the advanced Green function $G_{C}^{a}$ is the Hermitian conjugate of the retarded Green function $G_{C}^{r}$ of the conductor, and $\Gamma_{L,R}$ describe the
coupling between the conductor and the leads.
To compute the Green function of the conductor, we can write the expression of the retarded Green function of a system:
\begin{equation}\label{eq:4}
  G_{C}^{r}=[(E+i\eta)-H_{C}-\Sigma_{L}-\Sigma_{R}]^{-1},
\end{equation}
where $E$ is the quasiparticle energy measured with respect to the Fermi level $E_{f}$, and $\eta$ is a positive
infinitesimal number defining the ''retarded'' character of the Green function.
$H_{C}$ is the Hamiltonian matrix of the finite isolated conductor, $\Sigma_{L,R}$ are the retarded
self-energy terms due to the conductor coupling with the semi-infinite leads.
The self-energy terms are defined as:
\begin{equation}\label{eq:5}
  \Sigma_{L}=H_{LC}^{+}g_{L}H_{LC},
  \Sigma_{R}=H_{RC}^{+}g_{R}H_{RC},
\end{equation}
\begin{figure}[t]
  \centering
  \includegraphics[width=0.5\textwidth]{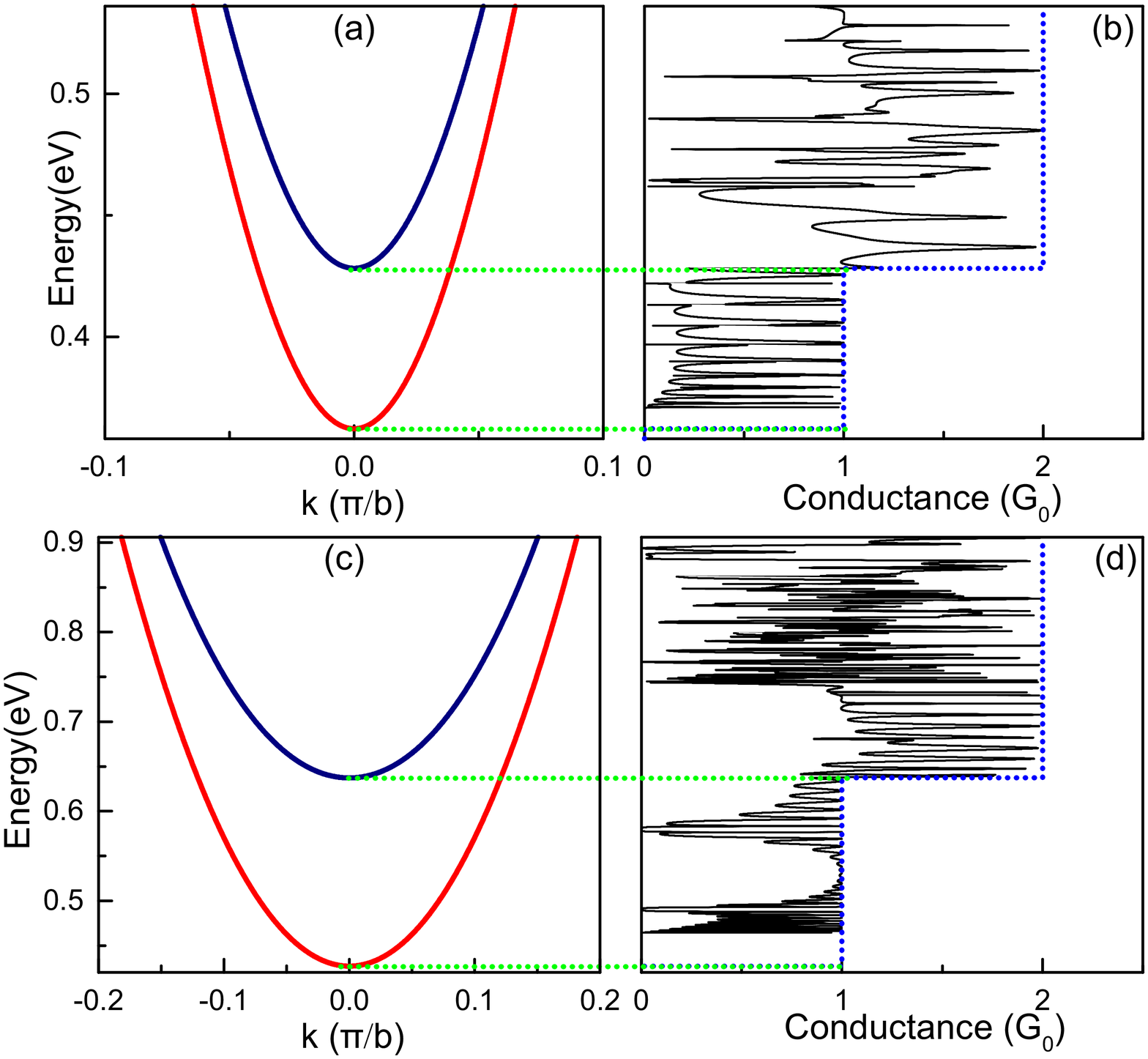}
  \caption{(Color online) Band structure of the lead and transport properties of (a) and (b) for armchair PNR, (c) and (d) for zigzag PNR. }\label{fig:2}
\end{figure}
where $H_{LC}$ and $H_{CR}$ representing the coupling matrices with non-zero elements only for adjacent lattices in the conductor and leads accounting for the nearest-neighbor TB approximation. $g_{L}$ and $g_{R}$ are the surface Green functions of the left and right semi-infinite leads. The self-energy term can be regarded as an
effective Hamiltonian that arises from the coupling of the conductor with leads. The key of the problem is how to obtain the surface
Green functions of the semi-infinite leads. Once the surface Green functions of the leads are known, the matrices $\Gamma_{L,R}$ can be easily obtained as:
\begin{equation}\label{eq:7}
  \Gamma_{L,R}=i[\Sigma_{L,R}^{r}-\Sigma_{L,R}^{a}],
\end{equation}
with the advanced self-energy $\Sigma_{L,R}^{r}$.

From Green's function, the local density of state (LDOS) at site i can be found:
\begin{equation}\label{eq:6}
 \rho_{i}=- \frac{1}{\Pi}Im[G_{i,i}],
\end{equation}
where $G_{i,i}$ is the matrix element of Green's function at site i.

To obtain the electron transport properties at finite temperature $(T)$, we use the non-zero temperature linear response formula:
\begin{equation}\label{eq:8}
  G(E_{F})=\frac{e^{2}}{\pi\hbar}{\int}T(E)F_{T}(E-E_{F})dE,
\end{equation}
where $F_{T}(E-E_{F})=-df(E)/dE$ is the thermal broadening function and $f(E)$ is
the Fermi-Dirac distribution function.

The magneticresistance (MR) is defined as:
\begin{equation}\label{eq:9}
  R_{M}(E_{F},B)\equiv[G(E_{F},0)-G(E_{F},B)]/G(E_{F},B).
\end{equation}
Here $G(E_{F},B)$ is the conductance of the system in a perpendicular magnetic field $B$ with a incident energy $E_{F}$.
\begin{figure}[b]
  \centering
  \includegraphics[width=0.5\textwidth]{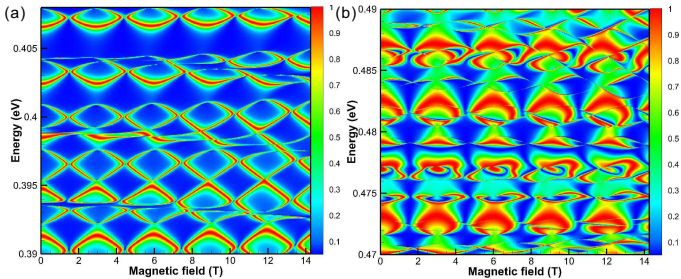}
  \caption{(Color online) The conductance of (a) armchair (b) zigzag PNR with structure parameters as $N_{L}=13$, $N_{M}=22$,  $N_{C}=22$,  $N_{T}=240$, $N_{H}=14$ at temperature $T=0$.  }\label{fig:3}
\end{figure}
\section{Aharonov-Bohm effect}
When we investigate Aharonov-Bohm effect in a nanoring, there is gauge freedom in the choice of
vector potential for a given magnetic field. The Hamiltonian is gauge invariant, which means that adding
the gradient of a scalar field to $\vec{A}$ changes the overall phase of the wave function by an amount
corresponding to the scalar field, physical properties are not influenced by the specific choice of
gauge. As we choose the Landau gauge in our calculations, we have $\vec{B}=\triangledown\times\vec{A}$ which is simply
the definition of vector potential and
\begin{equation}
\oint_{C}\vec{A}{\cdot}d{\vec{r}}=\int_{S}(\triangledown\times\vec{A}){\cdot}d{\vec{S}}=
\int_{S}\vec{B}\cdot{\vec{dS}}=\phi_{m}
\end{equation}
which is a consequence of Stokes theorem. $\phi_{m}$ is total magnetic flux through encircled by path $I$ and path $II$ shown in Fig.\ref{fig:1}.
When electron transmit through path $I$ and $II$ (Fig.\ref{fig:1}) in the presence of magnetic field and finally combine in
the right lead, the magnetic interference phase is $e^{i{\Delta}\phi}$,
\begin{align*}
{\Delta}\phi &=\frac{e}{\hbar}\left[\int_{C_{I}}\vec{A}(\vec{r})\cdot{d\vec{r}}-\int_{C_{II}}\vec{A}
(\vec{r})\cdot{d\vec{r}}\right]   \\
&=\frac{e}{\hbar}\oint_{C}\vec{A}(\vec{r})\cdot{d\vec{r}}   \\
&=\frac{e}{\hbar}\int_{S}\vec{B}\cdot{\vec{dS}}   \\
\end{align*}
here we don't discuss the normal interference phase when electron transmit through path $I$ and $II$ in the absence of magnetic field.
\par
The interference phase ${\Delta}\phi=\frac{e}{\hbar}B(S_{h}+S_{l})$, $S_{h}+S_{l}$
is the total area encircled by path $I$ and path $II$, $S_{h}$ is the area of the hole
in the nanoring, $S_{l}$ is the remain part encircled by the path on the phosphorene lattice.
$S_{l}$ is determined by the incident energy as for different incident energy the path is different. $\frac{e}{\hbar}BS_{h}$ is the non-local part of Aharonov-Bohm effect.
\par
The size of the rectangular PNRs are characterized by parameters shown in Fig.~\ref{fig:1}, the parameters are in
the unit of phosphorene lattice constant \textbf{a} or \textbf{b}. Normally, people can simulate relatively
small systems with recursive Green's function method, due to the cubic scaling of the computational burden associated with matrix inversion. In this work, we use recursive Green's function method, which cuts the whole system into many slices and the matrix inversion is calculated for each slice instead of the whole system. So this method enables the simulation of very long system and is preferred to studying quasi one dimensional systems, such as nanotubes and nanoribbons. Therefore, in this paper we consider PNRs which are narrow but relatively long. Two set of structure parameters are taken, the first set is $N_{L}=13$, $N_{M}=11$,  $N_{C}=11$,  $N_{T}=120$, $N_{H}=7$, the second set is $N_{L}=13$, $N_{M}=22$,  $N_{C}=22$,  $N_{T}=240$, $N_{H}=14$. The average area of the nanoring is given by $\bar{S}=(S_{inn}+S_{out})/2$, which is the average area of the inner ($S_{inn}$) and outer ($S_{out}$) rings. Then the average areas of the first kind of PNRs are $\bar{S}\approx359.4\ nm^{2}$, the average areas of another kind of PNRs are four times than that of the first kind. In the remain part, the calculations are based on the first kind of PNRs without specification as only one simulation result is based on the first kind of PNRs.
\par
\begin{figure}[t]
  \centering
  \includegraphics[width=0.5\textwidth]{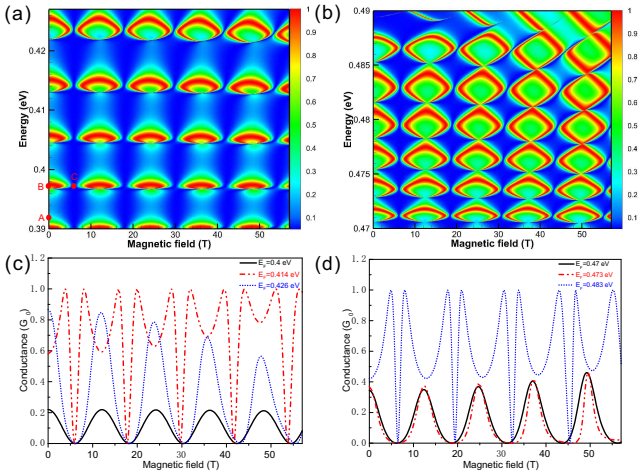}
  \caption{(Color online) The contour plot of conductance as a function of magnetic field and Fermi energy with temperature $T=0$ for (a) armchair nanoring,  (b) zigzag nanoring. The conductance with different incident energy with $T=0$ for (c) armchair nanoring and (d) zigzag nanoring. }\label{fig:4}
\end{figure}
First we calculate the band structure of the lead (semi infinite nanoribbons) and the conductance of the nanorings in the absence of magnetic fields at zero temperature. Fig.~\ref{fig:2}(a), (c) show the band dispersion of the lowest two conduction subbands of perfect armchair and zigzag ribbons, respectively. Unlike the features of graphene nanoribbons. Both armchair and zigzag PNRs possess finite band gaps. Accordingly the conductance of armchair/zigzag PNR is fully suppressed when the electron incident energy $E_{F}$ is below the bottom of the first subband, i.e., 0.37eV/0.47eV as shown in Fig.~\ref{fig:2}(b)/(d).  These low energy transmission-forbidden regions arise from the absence of propagating modes in the leads as we use a semi-infinite phosphorene nanoribbon as the lead, so when the incident energy is below the CBM of the lead or the lowest unoccupied molecular orbital (LUMO) of the nanoring, the transmission is forbidden.
\par
As the incident Fermi energy increases, we observe intensive oscillations caused by the Fabry-Perot resonant modes formed in the PNRs. Many valleys with zero or near zero conductance are observed arising from the absence of bound states in the central PNRs instead of lacking of propagating modes in the leads. The conductance also exhibits step-like behavior in agreement with the opening of new subbands.  Let us take the armchair PNR for example as shown in
Fig.~\ref{fig:2}(b). When $0.37\ eV<E_{F}<0.43\ eV$, the conductance of the ring can reach $G_{0}$, i.e., resonant conductance peaks. For larger incident energies in the range of $0.43\ eV<E_{F}<0.6\ eV$, the conductance oscillations become more complex, disordered, and the peaks approach $2G_{0}$ as the second subband start to contribute to the total conductance. Note that the main purpose of this work is investigate the phase coherent transport of the carriers in PNRs. It is preferable examine the energy region in which only one mode is engaged in electron transport. Hence, we only present the results with energy area corresponding to the first conduction subbands for both armchair and zigzag PNRs when we discuss AB oscillations and MR in PNRs.
\par
\begin{figure}[t]
  \centering
  \includegraphics[width=0.5\textwidth]{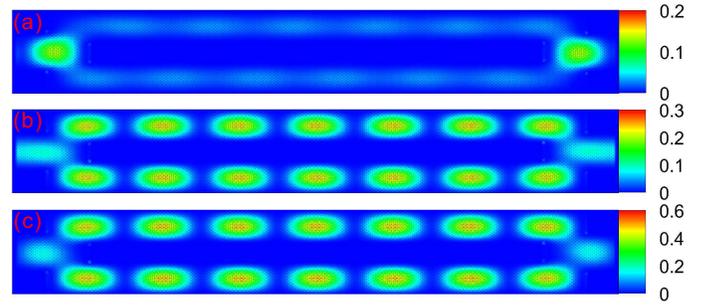}
  \caption{(Color online) LDOS of the nanoring of armchair nanoring. (a), (b), (c) correspond to the points marked A, B, C in Fig.~\ref{fig:4}(a). Here we use a
Gauss function to smear the contour plot.}\label{fig:5}
\end{figure}
The AB oscillations in mesoscopic rings are of particular interest and offering an elegant way to study phase-coherent electron transport properties. In the presence of a
perpendicular magnetic field \textbf{B}, electrons passing through either side of the PNR (path $I$ and path $II$ shown in Fig.~\ref{fig:1}) and this difference produces the phase modulation: ${\Delta}\phi=\frac{e}{\hbar}BS$.
Therefore, the transmission probability through the PNR exhibits periodic oscillations when varying the magnetic field with fixed period of $\Delta\bm{B}=2\pi\phi_{0}/S$.
\par
Next, we address that the conductance of a PNR does not only exhibit resonant behavior with incident Fermi energy due to the formation of Fabry-Perot modes but also oscillates with a perpendicular magnetic field arising from the formation of Landau levels and AB interference.
Fig.~\ref{fig:3} (a) and (b) show the contour plot of the conductance as a function of the incident Fermi energy and the magnetic field in armchair and zigzag PNRs respectively with structure parameters as $N_{L}=13$, $N_{M}=22$,  $N_{C}=22$, $N_{T}=240$, $N_{H}=14$. The area of these two nanoring are relatively large then the AB oscillation period $\Delta\bm{B}$ will be relatively small which can be realized experimentally. From the contour plot we can see that the conductance of the PNRs oscillates both with the incident energy and the magnetic field. The conductance oscillates with the incident energy because of resonant tunneling and oscillates with the magnetic field because of the AB effect. We find that the AB oscillation periods at different incident energies are slightly different. The reason for such a
difference is that for different incident energies the charge distributions in the arms of the central PNR are different and thus the effective areas encircled by path $I$ and path $II$ (Fig.~\ref{fig:1}) are different. In Fig.~\ref{fig:3}, the incident energy is in the interval only the first conduction subband of the bulk states take place. In this case, the PNR area can be approximated by $\bar{S}$ defined before which is about $1437.6\ nm^{2}$, then the oscillation period should be ${\Delta}B{\approx}2.88T$. The numerical results of ${\Delta}B$ shown in both Fig.~\ref{fig:3}(a) and Fig.~\ref{fig:3}(b) are about $3T$ which match well with the theoretical prediction. In the following part, we will illustrate this effect in more details.
\par
In Fig.~\ref{fig:4}(a), we plot the conductance of an armchair PNR as a function of the magnetic field strength \textbf{B} and the Fermi energy $E_{F}$ with structure parameters $N_{L}=13$, $N_{M}=11$,  $N_{C}=11$,  $N_{T}=120$, $N_{H}=7$ at zero temperature $T=0$. The conductance of armchair PNR oscillates periodically in magnetic field $B$. The period of $~12\ T$ is consistent with the expectation ($\Delta\bm{B}=2\pi\phi_{0}/\bar{S}$.) accounting for the area of our PNR as a manifestation of the AB effect. Importantly, the conductance peaks or valleys appear synchronously in magnetic field with varied $E_{F}$, since the phase modulation depends on the magnetic flux through PNR area rather than the incident energy. In such a small energy interval the effective PNR areas encircled by path $I$ and path $II$ all approximate to $\bar{S}$. Due to the contribution from both Fabry-Perot resonant and AB oscillations, the contour plot of the conductance exhibits beautiful fish scales. Fig.~\ref{fig:4}(b) is similar to Fig.~\ref{fig:4}(a) except for the PNR orientation, i.e., an zigzag PNR. In Figs.~\ref{fig:4}(c) and (d), we extract the
conductance of armchair and zigzag PNRs in varied magnetic fields at three different incident Fermi energies. The oscillation periods are slightly relate to the Fermi energies, in Fig.~\ref{fig:4}(c) and (d) we can see that the conductance for different incident energy are not exactly aligned due to the same reason as we discussed for Fig.~\ref{fig:3}. We stress in advance that this small variation can hardly suppress the AB effect induced giant MR especially for the first MR peak as we will discuss in the next section. The Fermi energy can also affect the conductance maxima via the density modulation. In addition at certain Fermi energies, double peaks in the conductance verse magnetic field plot are observed. These double peaks come out when the AB destructive interference regions cross the resonant peaks with high density of bound states.
The differences between Fig.~\ref{fig:4}(a) (c) and Fig.~\ref{fig:4}(b) (d) originate from the anisotropic resonant tunneling behaviors in armchair and zigzag PNRs which have been shown in Fig.~\ref{fig:2} above.

\begin{figure}[t]
  \centering
  \includegraphics[width=0.5\textwidth]{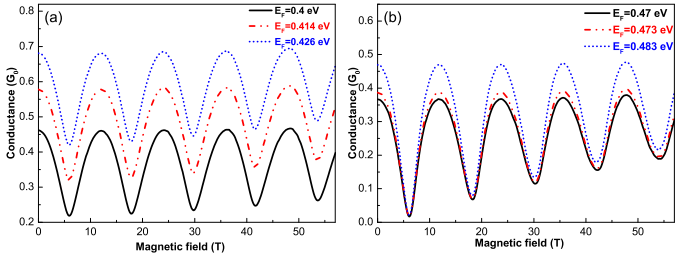}
  \caption{(Color online) The relation beteen the conductance and magnetic field of armchair (a) and zigzag (b) PNR with different incident energy at room temperature.  }\label{fig:6}
\end{figure}
To clarify the origin of the resonant maxima and different types of the minima, we plot the LDOS of armchair PNRs in Fig.~\ref{fig:5} (a) - (c), corresponding to the marks $A$, $B$, $C$ shown in Fig.~\ref{fig:4}(a). It is helpful to distinguish the Fabry-Perot interferences and AB interference effects in the PNRs. We must
point out that the LDOS shown in Fig.~\ref{fig:5} are smeared by Gauss function. In the upper and lower bridges
of the armchair PNR ring, Fabry-perot modes can be formed as a result of the quantum interferences between electron
waves moving forward and backward. The presence or absence of these quasi bound states is determined by Fermi
energy and ring size $L_{T}$. Heuristically, the quantization conditions for the
bound states are typically given by $n \cdot \lambda = L_{T}$, where $n$ is an integer and $\lambda$ is the electron wave
length satisfying the relation ship $E_{F}=hv_{F}/\lambda$. The density of states develops peaks at $E =nhv_{F}/L_{T}$. In Fig.~\ref{fig:5}(a), the Fermi energy is set to 0.394 eV and the magnetic field $B=0$, which corresponds to the fully blocked case as indicated by mark A in Fig~\ref{fig:4} (a). The LDOS are mostly
concentrated in the left and right leads, while few bound states are formed in the two ring bridges. So electrons
can hardly propagate through the PNR. We also plot the spatial distribution of LDOS corresponding to the
conduction peak in Fig~\ref{fig:5}(b), with $E_{F}$= 0.397 eV, and  B=0 (see mark B in Fig.~\ref{fig:4}(a)). Much more bound states are formed in
the each side of the PNR, that can assist electron transmission and finally give rise to a conductance peak.
Interestingly, with the same $E_{F}$ level of 0.397 eV, but increase the magnetic field from 0 T to 6.5 T as indicated by
mark C in Fig~\ref{fig:4}(a), the electron transmission is fully suppressed while the bound states in the bridges are
mainly preserved as shown in Fig.~\ref{fig:5}(c). This conductance dip comes from the destructive
interferences at the exit interconnection of two paths, i.e., the AB effect. Electrons from upper or lower paths gain different phase shift arising from the magnetic flux in the PNR. We therefore confirm the AB effect in this PNR.

\begin{figure}[t]
  \centering
  \includegraphics[width=0.5\textwidth]{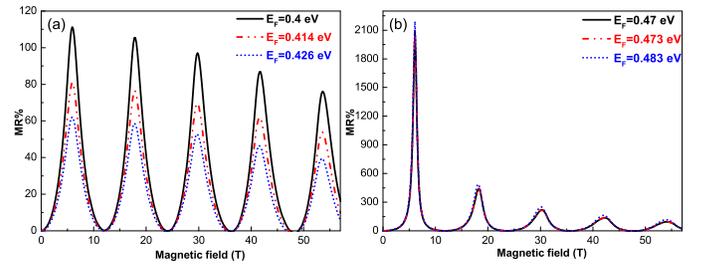}
  \caption{(Color online) Magnetic resistance at room temperature of (a) armchair, (b) zigzag PNR with three different incident energys.  }\label{fig:7}
\end{figure}
\section{Magnetoresistance of phosphorene nanorings}
Next we explore how a finite temperature affects the conductance of the PNRs. A thermal broadening function is taken into consideration in the calculation at non-zero temperature.
At room temperature (298$K$), the oscillation behavior is quite different from that at zero temperature as shown in Fig.~\ref{fig:4}. The double peaks of conductance at zero temperature disappear and the oscillation amplitudes are reduced. The
conductance of the armchair PNR oscillates in between 0.2 $G_{0}$ and 0.7 $G_{0}$, see Fig.~\ref{fig:6}(a).
The oscillation minima of different Fermi energy almost keep steady in the armchair PNR,
while in the zigzag PNR, the minima increase with $B$ (Fig.~\ref{fig:6}(b)).
\par
Finally, the MR of both armchair and zigzag nanoring are calculated at room temperature, see Fig.~\ref{fig:7}.
It is shown that the conductance of armchair nanoring exhibits clear AB oscillations (Fig.~\ref{fig:7}(a)),
the period of which also matches well with the expression $\Delta\bm{B}=2\pi\phi_{0}/\bar{S}$. There is a closely connection
between MR and Fermi energy, i.e., higher Fermi energy gives rise to lower MR. The order of the MR for this armchair nanoring is about hundred percent, which is much smaller than that of an armchair graphene nanoring\cite{Nguyen}.
While in zigzag nanoring, see Fig.~\ref{fig:7}(b), we find that the MR can be as giant as two thousand percent which is much greater
than that in armchair nanoring which coincidence with the anisotropy electron property of
phosphorene, the MR decay rapidly with magnetic field, and the MR with different Fermi energy are almost the same.
\section{SUMMARY}
In this paper, we theoretically demonstrate the AB effect in monolayer PNRs utilizing TB method and recursive Green's function method. Our numerical results show that the conductance of PNRs oscillates dramatically with the incident Fermi energy and the perpendicular magnetic field. The complex oscillating behaviors arises from hybrid effects of formation of Fabry-Perot modes, formation of Landau levels and the AB interference. The AB oscillation period is dominated by the effect area of the PNR and slightly affected by the incident Fermi energy. By limiting the incident Fermi energy lower than the bottom of the second subband, the AB effect become more pronounced than other effects, leading to a giant MR in PNR. The MR is highly anisotropic depending on the PNR orientation, i.e., the maximum MR of the zigzag PNR is one order of magnitude larger than that of the counterpart armchair PNR. This investigation sheds new light on constructing phosphorene based nanoelectronic devices.

\bibliographystyle{apsrev4-1}
\bibliography{ring}

\end{document}